\documentstyle[epsfig,12pt]{article}
\newcommand{\be}{\begin{eqnarray}}
\newcommand{\ee}{\end{eqnarray}}

\def\lsim{\mathrel{\rlap{\lower4pt\hbox{\hskip1pt$\sim$}}
    \raise1pt\hbox{$<$}}}               
\def\gsim{\mathrel{\rlap{\lower4pt\hbox{\hskip1pt$\sim$}}
    \raise1pt\hbox{$>$}}}               

\def\dslash{\slash \hskip-5pt}  

\begin{document}

\rightline{\Large{Preprint RM3-TH/01-6}}

\vspace{2cm}

\begin{center}

\LARGE{Relativistic quark models\footnote{To appear in the Proceedings of the International Conference on {\em The Physics of Excited Nucleons} ($N^* ~ 2001$), Mainz (Germany), March 7-10, 2001, World Scientific Publishing (Singapore), in press.}}

\vspace{1cm}

\large{Silvano Simula}

\vspace{0.5cm}

\normalsize{Istituto Nazionale di Fisica Nucleare, Sezione Roma III\\ Via della Vasca Navale 84, I-00146 Roma, Italy}

\end{center}

\vspace{2cm}

\begin{abstract}

\noindent The application of relativistic constituent quark models to the evaluation of the electromagnetic properties of the nucleon and its resonances is addressed. The role of the pair creation process in the Feynmann triangle diagram is discussed and the importance both of choosing the light-front formalism and of using a Breit frame where the {\em plus} component of the four-momentum transfer is vanishing, is stressed. The nucleon elastic form factors are calculated free of spurious effects related to the loss of rotational covariance. The effect of a finite constituent size is considered and the parameters describing the constituent form factors are fixed using only the nucleon data up to $Q^2 \simeq 1 ~ (GeV/c)^2$; a constituent charge radius of $\simeq 0.45 ~ fm$ is obtained in this way. Above $Q^2 \simeq 1 ~ (GeV/c)^2$ our parameter-free predictions are in good agreement with the experimental data, showing that soft physics may be dominant up to $Q^2 \sim 10 ~ (GeV/c)^2$. Our new results for the longitudinal and transverse helicity amplitudes of the $N - P_{11}(1440)$ transition are presented.

\end{abstract}

\newpage

\pagestyle{plain}

\section{Introduction}

\indent The electromagnetic (e.m.) properties of the nucleon and its resonances contain important pieces of information on the internal structure of the nucleon, and therefore an extensive experimental program aimed at their determination is currently undergoing and planned at several facilities around the world. As it is well known, a successful phenomenological model for the description of the hadron structure is based on the concept of constituent quarks ($CQ$'s) as effective degrees of freedom in hadrons. The $CQ$ model assumes that the degrees of freedom associated to gluons and $q \bar{q}$ pairs are frozen, while their effects are hidden in the $CQ$ mass and (possible) size as well as in the effective interaction among $CQ$'s. Originally the $CQ$ model was formulated in a non-relativistic way, and indeed its first success was the reproduction of the mass spectra of mesons with two heavy quarks \cite{Bali}. However, the non-relativistic $CQ$ model is unexpectedly successful also in the light-quark sector. As a matter of fact it suffices to remind that the simple assumption of an $SU(6)$ symmetry of the nucleon wave function leads immediately to several predictions, which are in qualitative agreement with experimental data; among them we mention: ~ i) the neutron electric form factor $G_E^n(Q^2)$ is vanishing; ~ ii) the neutron to proton ratio of magnetic form factors $G_M^n(Q^2) / G_M^p(Q^2)$ is independent of $Q^2$ and equal to $- 2 / 3$; ~ iii) all the radiative transitions among members of the $[56, 0^+]$ and $[70, 1^-]$ $SU(6)$-multiplets can be written in terms of only three independent parameters \cite{Burkert}.

\indent The non-relativistic $CQ$ model has several flaws, which were clearly pointed out in \cite{Isgur}. Here we limit ourselves to mention the most relevant inconsistency, namely the fact that the average momentum of a light $CQ$ in a hadron is definitely larger than its mass (see, e.g., the $CQ$ momentum distribution in the nucleon and in the most prominent electroproduced resonances reported in  \cite{CAR99}). Therefore, relativized versions of the $CQ$ effective Hamiltonian have been proposed in the literature, trying mainly to take into account the relativistic form of the kinetic energy and the relativistic smearing of the $CQ$ positions, as it is the case of the one-gluon-exchange ($OGE$) model \cite{Isgur} and of the chiral model based on Goldstone-boson-exchange ($GBE$) mechanism \cite{Glozmann}. Nevertheless, for both relativized models the proton size turns out to be much smaller ($\simeq 0.3 ~ fm$) than its experimental charge radius ($\simeq 0.8 ~ fm$), showing the need for a proper treatment of relativistic effects in the evaluation of the nucleon e.m. properties even close to the photon point (see \cite{Fabio}).

\section{Relativistic constituent quark models}

\indent There are three main ways to include the interaction in the generators of the Poincar\'e group and to maximize at the same time the number of kinematical (interaction-free) generators. The three relativistic forms are: ~ i) the instant form, where the interaction is present in the time component $P^0$ of the four-momentum and in the Lorentz boost operators; ~ ii) the point form ($PF$), where all the components of the four-momentum operator depend on the interaction, and ~ iii) the light-front ($LF$) form, in which the interaction appears in the "minus" component of the four-momentum ($P^- \equiv P^0 - P^z$) and in the transverse rotations around the $x$ and $y$ axes (with the null-plane being defined by $x^+ \equiv t + z = 0$). Since in the instant form it is not possible to define kinematical boost operators, in what follows we will limit ourselves only  to the $PF$ and to the $LF$ formalism.

\indent In recent years relativistic $CQ$ models employing the two above-mentioned forms of the dynamics have been used for the investigation of hadron e.m. properties. In such applications the one-body approximation for the e.m. current has been widely considered, because it allows to address the important issue to know to what extent the e.m. properties of a composite system can be understood in terms of the e.m. properties of its constituents. The one-body approximation can be generally written as
 \be
      J^{\mu} \approx J_1^{\mu} = \sum_{j = 1}^N \left[ f_1^{(j)}(Q^2) 
      \gamma^{\mu} + f_2^{(j)}(Q^2) {i \sigma^{\mu \nu} q_{\nu} \over 2m_j} 
      \right]
      \label{eq:one-body}
 \ee
where $Q^2 \equiv - q \cdot q$ is the squared four-momentum transfer and $f_{1(2)}^{(j)}(Q^2)$ are Dirac ({Pauli) form factors of the $j$-th constituent quark with mass $m_j$. While the exact current $J^{\mu}$ is Poincar\'e covariant in all the three forms, the approximate current $J_1^{\mu}$ is covariant only in the $PF$.

\indent However, if a connection between relativistic quantum models and covariant quantum field theories is searched, a minimal requirement is the matching with the Feynmann one-loop triangle diagram. As it is well known \cite{Zgraph}, the latter is given by the sum of two terms, namely the spectator-pole and the pair creation diagram (the $Z$-graph). The spectator term corresponds to a one-body process, while the $Z$-graph is a many-body process. In the $PF$ the two terms are separately covariant, while in the $LF$ formalism only their sum is covariant. This means that only in the $LF$ formalism one may find a reference frame where the $Z$-graph is suppressed. As a matter of fact, it is well known \cite{Zgraph} that such a reference frame is the Breit frame where $q^+ \equiv q^0 + q^z = 0$.

\indent In the $PF$ the contribution of the one-body current (\ref{eq:one-body}) does not correspond to the spectator term of the triangle diagram, so that in order to match the latter one needs to include explicitly the $Z$-graph. A manifestation of the mismatch of the spectator term with the one-body $PF$ approximation is the fact that the squared four-momentum transfer seen by the constituent $\bar{Q}^2$ differs from $Q^2$ (see \cite{Klink}). On the contrary in the $LF$ formalism the use of a Breit frame where $q^+ = 0$ makes the one-body contribution equivalent to the spectator term \cite{Zgraph}, while at the same time the $Z$-graph is suppressed. Note that $\bar{Q}^2 = Q^2$ only when $q^+ = 0$. However, performing the limit $q^+ \to 0$ of the $Z$-graph requires some care. As properly pointed out in \cite{Frederico}, the longitudinal zero-momentum mode may contribute to few matrix elements of the current components even when $q^+ = 0$ and such contributions turn out to be essential in order to recover the full covariance of the calculations. Therefore, the strategy to evaluate the form factors encoded in the triangle diagram is to identify the subset of amplitudes characterized by the matching of the triangle diagram with the one-body $LF$ approximation at $q^+ = 0$. To carry out this strategy a useful formalism is the covariant $LF$ approach developed in \cite{Karmanov}. There, it has been argued that the loss of covariance for an approximate current implies the dependence of its matrix elements upon the choice of the null four-vector $\omega$ defining the spin-quantization axis. The particular direction defined by $\omega$ is irrelevant only if the current satisfies covariance. Therefore, the covariant decomposition of an approximate current can still be done provided the four-vector $\omega$ is explicitly considered in the construction of the relevant covariant structures. If we indicate by $\Gamma^{\mu; \alpha ... \beta ...}$ the amplitude tensor corresponding to the triangle diagram, where $\mu$ is the index for the current and $\alpha ...$ ($\beta ...$) are the Lorentz indices describing the spinorial structure of initial (final) bound-state vertex, one has (at $q^+ \equiv \omega \cdot q = 0$)
 \be
       \Gamma_{LF}^{\mu; \alpha ... \beta ...}(\omega) = \Gamma^{\mu; \alpha 
       ... \beta ...} + B^{\mu; \alpha ... \beta ...} (\omega)
       \label{eq:Gamma}
 \ee
where $\Gamma_{LF}^{\mu; \alpha ... \beta ...}$ corresponds to a given approximate $LF$ current, like the one-body approximation (\ref{eq:one-body}), and $B^{\mu; \alpha ... \beta ...} (\omega)$ is the tensor depending on $\omega$ and describing the mismatch between the triangle diagram and the given $LF$ approximation. Note that $\omega$ is proportional to the difference between the Feynmann and $LF$ momenta of the struck constituent, viz. $p - p_{LF} \propto \omega$, where $p$ is the constituent (off-mass-shell) momentum entering the triangle diagram (i.e., $p^2 \neq m^2$), while $p_{LF}$ is by definition constructed on-mass-shell (i.e., $p_{LF}^2 = m^2$). In Eq. (\ref{eq:Gamma}) the tensor $B^{\mu; \alpha ... \beta ...} (\omega)$ is responsible for the so-called violations of the angular condition, which have been explicitly estimated in \cite{CAR_rho} and \cite{CAR_Delta} in case of the $\rho$-meson and the $N - \Delta(1232)$ transitions, respectively, and also in \cite{KM} and \cite{Fabio}(b) in case of the nucleon elastic form factors.

\section{Nucleon elastic form factors and the $N - P_{11}(1440)$ transition}

\indent Let us now consider explicitly the case of the e.m. transition to a nucleon resonance with same quantum number $J^P = 1/2^+$ of the nucleon and indicate by $\bar{u}(p^* \nu') \Gamma^{\mu} u(p \nu)$ the matrix elements of the e.m. current corresponding to the triangle diagram for an initial (final) helicity equal to $\nu$ ($\nu'$). One has
 \be
       \Gamma^{\mu} = F_1^{N^*}(Q^2) \left[ \gamma^{\mu} + q^{\mu} {M^* - M 
       \over Q^2} \right] + F_2^{N^*}(Q^2) {i \sigma^{\mu \rho} q_{\rho} 
       \over M^* + M}
       \label{eq:triangle}
 \ee
where $p$ and $p^*$ are the four-momentum of the nucleon and its resonance with mass $M$ and $M^*$, respectively, and $F_j^{N^*}(Q^2)$ (with $j = 1, 2$) are the physical form factors. Following Eq. (\ref{eq:Gamma}) the $LF$ approximation can be written as: $\Gamma_{LF}^{\mu} = \Gamma^{\mu} + B^{\mu}(\omega) / (\omega \cdot P) $ with
 \be
       B^{\mu}(\omega) & = & B_1^{N^*}(Q^2) \left[ \dslash{\omega} - {2 
       \omega \cdot P \over (M^* + M) (1 + \eta)} \right] \left[ P^{\mu} + 
       q^{\mu} {M^{*2} - M^2 \over 2Q^2} \right] \nonumber \\
       & + & B_2^{N^*}(Q^2) \omega^{\mu} + B_3^{N^*}(Q^2) \omega^{\mu} 
       {\dslash{\omega} \over \omega \cdot P} + (M^* - M) B_4^{N^*}(Q^2) 
       q^{\mu} {\dslash{\omega} 
       \over \omega \cdot P} \nonumber \\
       & + & (M^* - M) B_5^{N^*}(Q^2) i \sigma^{\mu \rho} \omega_{\rho} ~ ,
      \label{eq:B_LF}
 \ee
where $\eta \equiv Q^2 / (M^* + M)^2$, $P = (p + p^*) / 2$ and $B_j^{N^*}(Q^2)$ (with $j = 1, ..., 5$) are (unphysical) spurious form factors. Note that, while $\Gamma \cdot q = 0$, one has $\Gamma_{LF} \cdot q \neq 0$ because $B \cdot q\neq 0$. However, in the elastic case ($M^* = M$) the structures proportional to $B_4^{N^*}(Q^2)$ and $B_5^{N^*}(Q^2)$ are absent thanks to time reversal symmetry, so that one gets $\Gamma_{LF} \cdot q = 0$. Generalizing the results obtained in \cite{Fabio}(b) and adopting our standard Breit frame where $q = (q_x, q_y, q^+, q^-) = (Q, 0, 0, 0)$, one gets
 \be
       G_E^{N ^*}(Q^2) & \equiv & F_1^{N^*}(Q^2) - \eta F_2^{N^*}(Q^2) = 
       \mbox{Tr}\left\{ {\Gamma_{LF}^+ \over 4P^+} \right\} - {iQ \over M^* 
       + M} \mbox{Tr}\left\{ {\Gamma_{LF}^+ \over 4P^+} \sigma_y \right\} 
       \nonumber \\
       G_M^{N^*}(Q^2) & \equiv & F_1^{N^*}(Q^2) + F_2^{N^*}(Q^2) = - {i 
       \over 2Q} \mbox{Tr}\left\{\Gamma_{LF}^y \sigma_z \right\}
       \label{eq:GEGM}
 \ee
with $\sigma_y$ and $\sigma_z$ being Pauli matrices. Therefore, the charge form factor $G_E^{N^*}(Q^2)$ can be extracted from the matrix elements of the {\em plus} component of the current, while the magnetic form factor $G_M^{N^*}(Q^2)$ has to be determined from the $y$-component, as already obtained in the elastic case in Ref. \cite{Fabio}(b).

\begin{figure}[htb]

\begin{center}

\epsfxsize=12cm 

\epsfbox{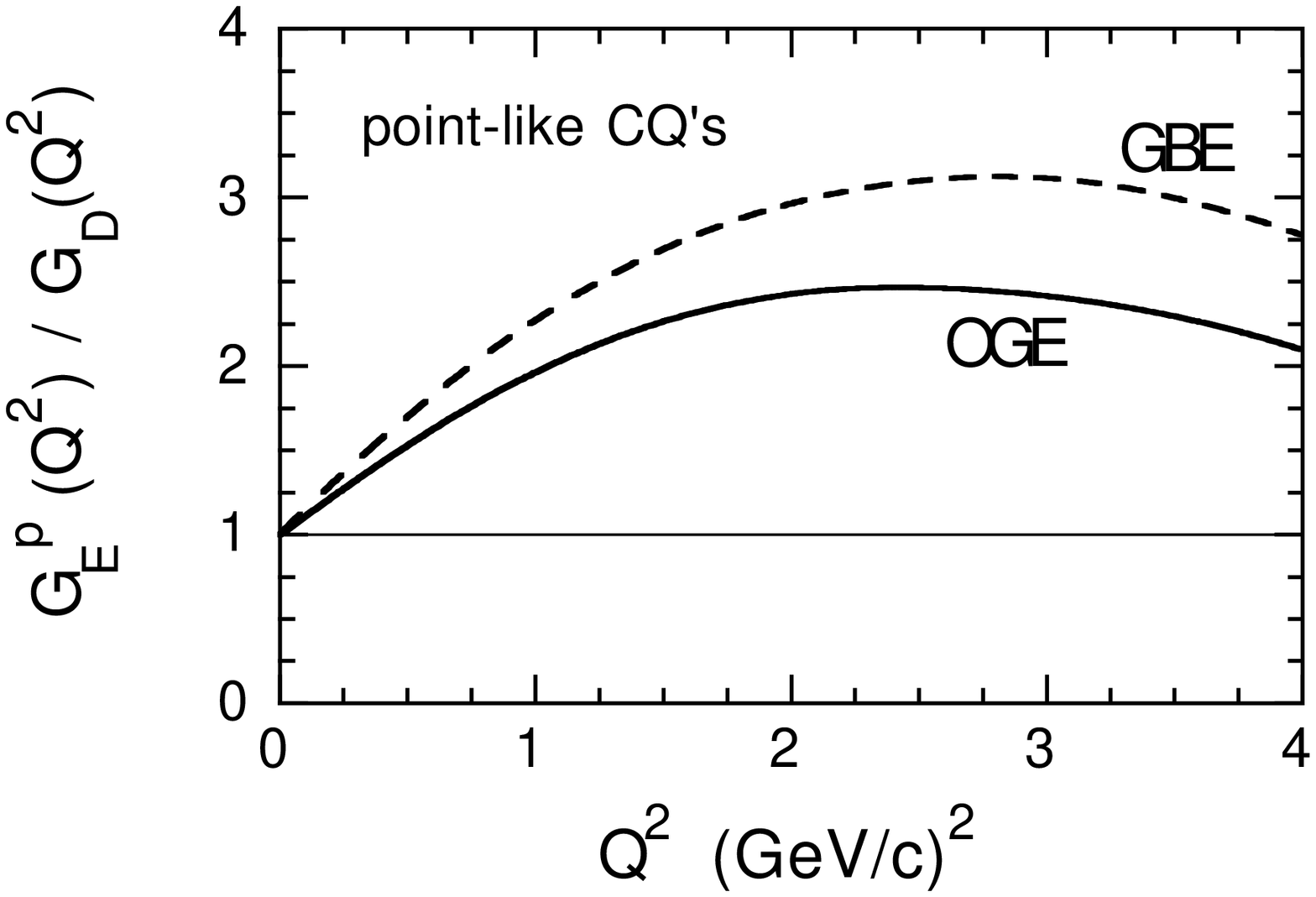}

\end{center}

{\small{{\bf Figure 1.} The proton charge form factor $G_E^p(Q^2)$, divided by the standard dipole form factor $G_D(Q^2) = 1 / (1 + Q^2 / 0.71)^2$, versus $Q^2$. The solid and dashed lines correspond to our $LF$ result (\protect\ref{eq:GEGM}) obtained assuming point-like $CQ$'s in Eq. (\protect\ref{eq:one-body}) and adopting the nucleon wave functions corresponding to the $OGE$ \protect\cite{Isgur}(b) and $GBE$ \protect\cite{Glozmann} models, respectively.}}

\end{figure}

\indent In Fig. 1 we have reported the results obtained for the proton charge form factor $G_E^p(Q^2)$ assuming point-like $CQ$'s in Eq. (\ref{eq:one-body}) and adopting the nucleon wave functions corresponding to the $OGE$ and $GBE$ models which provides a proton size of $\simeq 0.3 ~ fm$. It can clearly be seen that the introduction of boost effects increase remarkably the calculated proton charge radius (up to $\simeq 0.6 ~ fm$), but this is not sufficient to get agreement with its experimental value. Therefore, we now consider the effects of a finite $CQ$ size by introducing $CQ$ form factors in Eq. (\ref{eq:one-body}). The functional forms of $f_{1(2)}^{(j)}(Q^2)$ (with $j = u, d$) are the same as those adopted in \cite{CAR_N}; however, in this work the parameters are fixed by the requirement of reproducing the nucleon and pion data of \cite{CAR_N} only up to $Q^2 \simeq 1 ~ (GeV/c)^2$, i.e. up to the scale of chiral symmetry breaking. We stress that the basic difference with \cite{CAR_N} is that in this work by means of Eq. (\ref{eq:GEGM}) the nucleon form factors are evaluated free of the spurious effects arising from the loss of rotational covariance. The $CQ$ form factors, obtained from our least-$\chi^2$ procedure, provide a $CQ$ charge radius equal to $\simeq 0.45 ~ fm$, representing a clear improvement with respect to the findings of \cite{CAR_N}.

\begin{figure}[htb]

\begin{center}

\epsfxsize=15cm 

\epsfbox{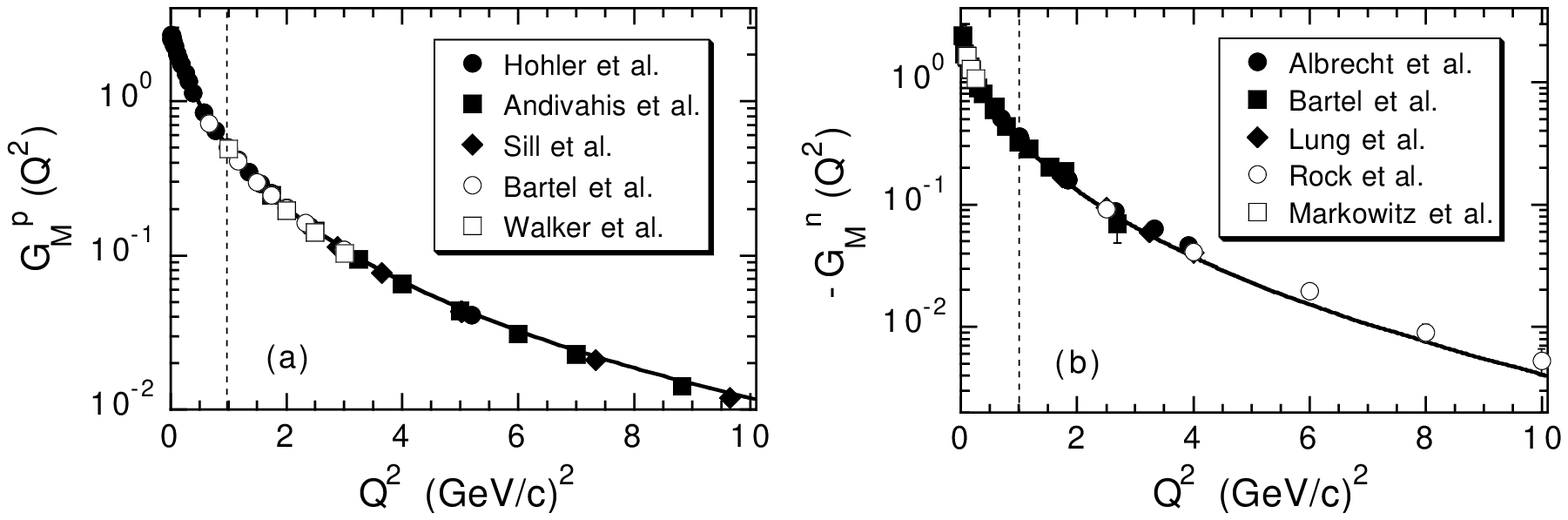}

\end{center}

{\small{{\bf Figure 2.} Proton (a) and neutron (b) magnetic form factors versus $Q^2$. The solid lines are our $LF$ results (\ref{eq:GEGM}) obtained adopting the $OGE$ model and including $CQ$ form factors in the one-body approximation (\ref{eq:one-body}). The dotted vertical lines correspond to $Q^2 = 1 ~ (GeV/c)^2$, which limits the range of values of $Q^2$ used for fixing the parameters of the $CQ$ form factors (see text). The data set is the same as in \protect\cite{CAR_N}.}}

\end{figure}

\indent Our results for all the nucleon form factors are reported in Figs. 2 and 3 in an extended range of values of $Q^2$. Note that above $Q^2 \simeq 1 ~ (GeV/c)^2$ our results should be considered as parameter-free predictions. It can be seen that a qualitative agreement with the data holds in a wide range of values of $Q^2$, and this is a clear indication that soft physics may play an important (even dominant) role also above the scale of chiral symmetry breaking, at least up to $Q^2 \simeq 10 ~ (GeV/c)^2$. Note that the $Q^2$-behavior obtained for $G_E^n(Q^2)$ [see Fig. 3(b)] is in overall agreement with available experimental data, but the neutron charge radius is remarkably underestimated (by $\simeq 50 \%$). Such a failure might be ascribed to the lack of explicit many-body currents in our present calculations.

\begin{figure}[htb]

\begin{center}

\epsfxsize=15cm 

\epsfbox{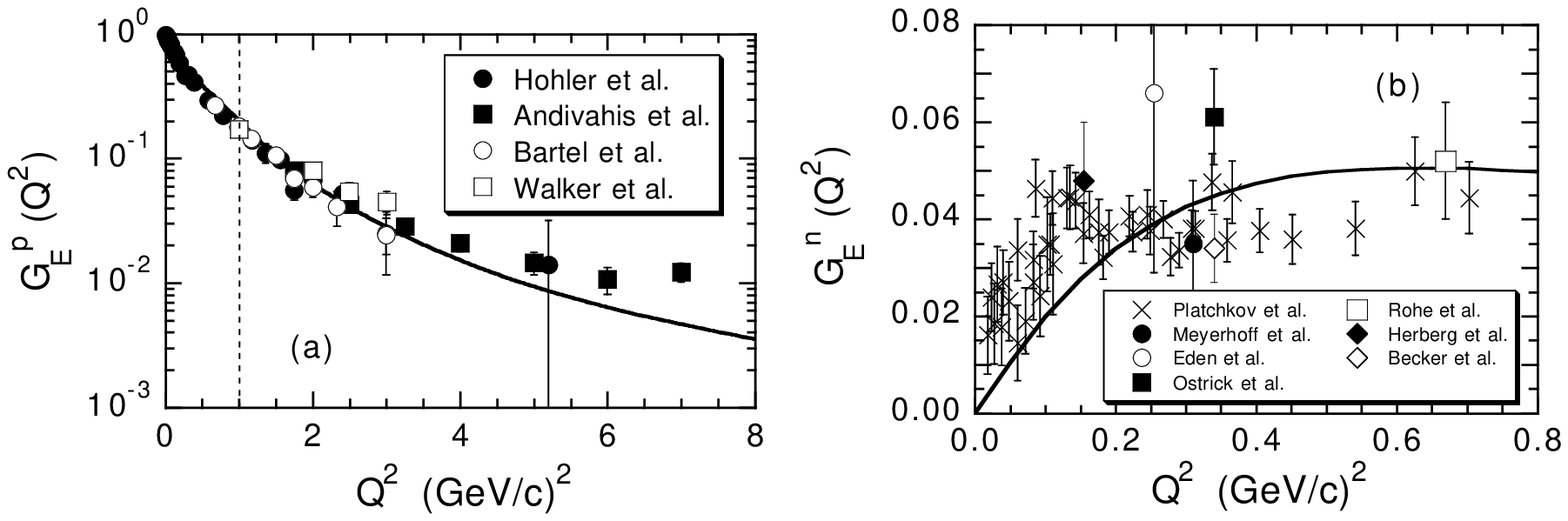}

\end{center}

{\small{{\bf Figure 3.} The same as in Fig. 2, but for the nucleon electric form factors. The data are from \protect\cite{CAR_N} (a) and \protect\cite{GEn} (b).}}

\end{figure}

\indent In Fig. 4 our results for the electric to magnetic proton ratio $\mu_p G_E^p(Q^2) / G_M^p(Q^2)$, where $\mu_p \equiv G_M^p(Q^2 = 0)$, are presented and compared with experimental data, including also the recent results from $JLAB$ \cite{JLAB}. It can be seen that above $Q^2 \simeq 1 ~ (GeV/c)^2$ our approach predicts a sharp suppression from the dipole law in nice agreement with the $JLAB$ data. Moreover, for $Q^2 \gsim 3 ~ (GeV/c)^2$ there is an interesting sensitivity to the inclusion of $CQ$ form factors, which could be further checked by future experiments.

\begin{figure}[htb]

\begin{center}

\epsfxsize=12cm 

\epsfbox{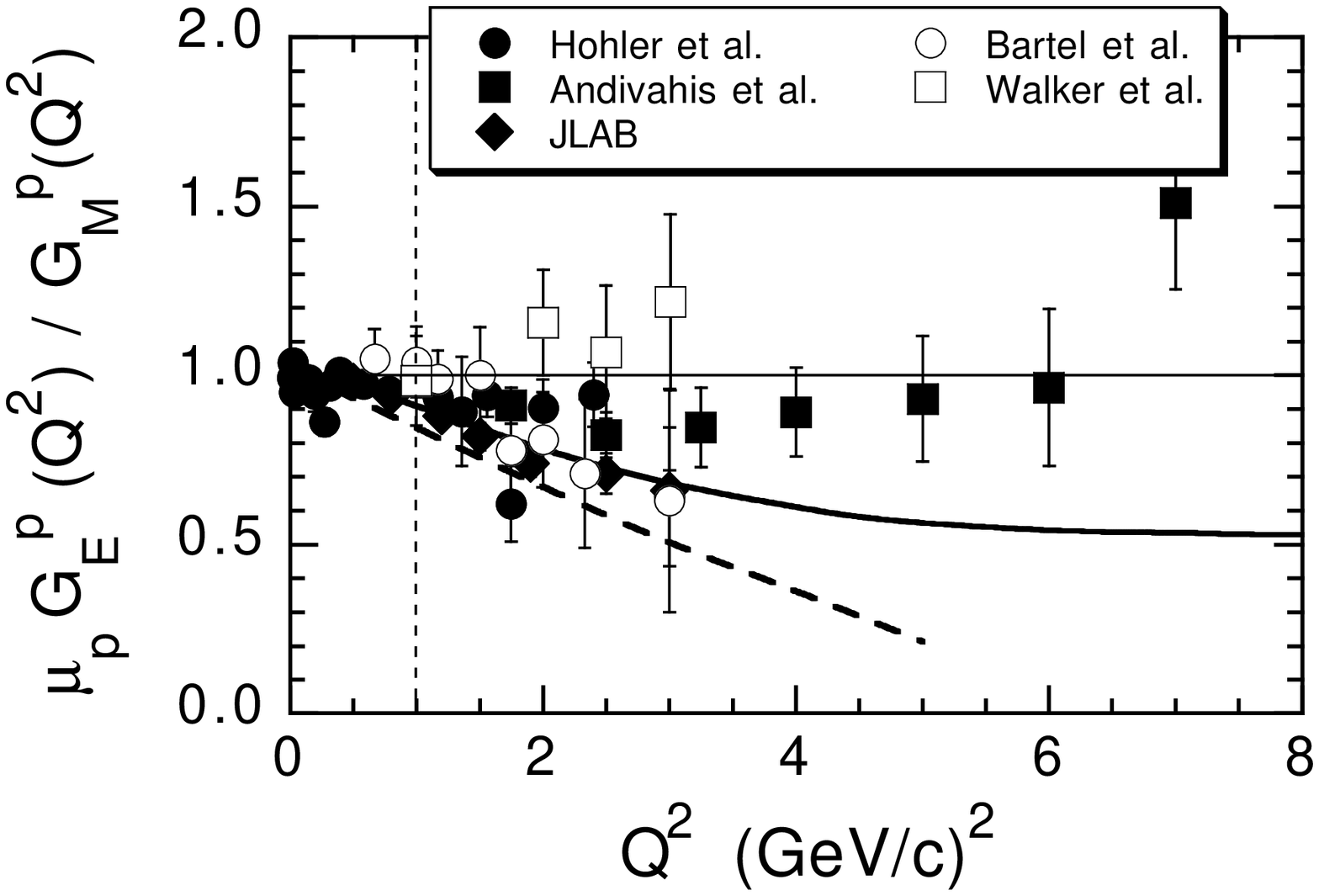}

\end{center}

{\small{{\bf Figure 4.} Electric to magnetic ratio $\mu_p G_E^p(Q^2) / G_M^p(Q^2)$ versus $Q^2$. The solid line is the $LF$ result including $CQ$ form factors, while the dashed line is taken from Ref. \protect\cite{Fabio}(b) and represents our $LF$ calculation assuming point-like $CQ$'s. The dotted vertical line has the same meaning as in Fig. 2. The data set is from \protect\cite{CAR_N} and \protect\cite{JLAB}.} }

\end{figure}

\indent We have applied Eq. (\ref{eq:GEGM}) to the calculation of the helicity amplitudes of the $N - P_{11}(1440)$ transition, which is of great interest because of a possible presence of a hybrid component in the Roper [$P_{11}(1440)$] resonance (see \cite{Li}). Our predictions are reported in Fig. 5 and compared with both the scarce experimental data and few theoretical predictions of other approaches. It can be seen that the boost effects reduce significantly both the transverse $A_{1/2} \propto G_M(Q^2)$ and the longitudinal $S_{1/2} \propto G_E(Q^2) / Q^2$ amplitudes (see \cite{CAR_Roper} for their precise definitions). However, the fastest fall-off is exhibited by the hybrid $q^3G$ model. Finally, note that our $LF$ result strongly underestimates the $PDG$ result for $A_{1/2}$ at the photon point.

\begin{figure}[htb]

\begin{center}

\epsfxsize=15cm 

\epsfbox{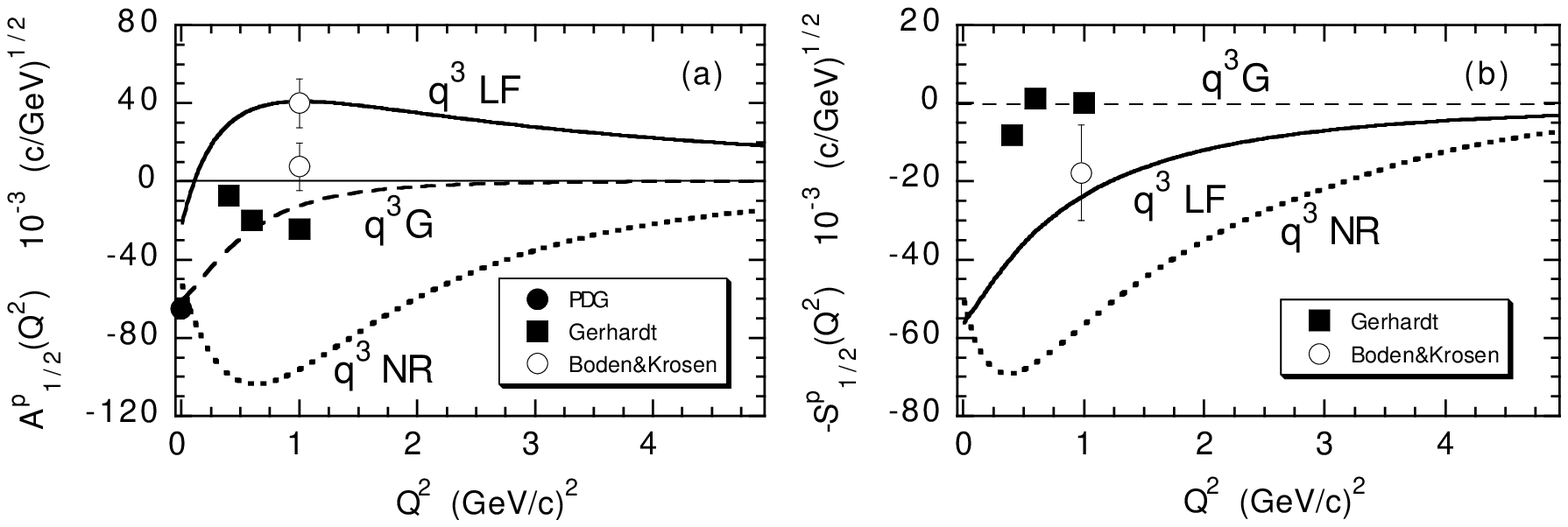}

\end{center}

{\small{{\bf Figure 5.} Transverse (a) and longitudinal (b) helicity amplitudes for the $p - P_{11}(1440)$ transition. The solid lines correspond to our $LF$ prediction, while the dotted and dashed lines are the result of the non-relativistic $q^3$ and of the hybrid $q^3G$ models obtained in \protect\cite{Li}. The meaning of the markers is the same as in \protect\cite{CAR_Roper}.}}

\end{figure}

\section{Conclusions}

\indent The application of relativistic $CQ$ models to the evaluation of the e.m. properties of the nucleon and its resonances has been addressed. The role of the pair creation process in the Feynmann triangle diagram has been discussed and the importance both of choosing the $LF$ formalism and of using a Breit frame where $q^+ = 0$, has been stressed. The nucleon elastic form factors have been calculated free of the spurious effects related to the loss of rotational covariance. The effect of a finite $CQ$ size has been considered and the parameters describing the $CQ$ form factors have been fixed using only the nucleon data up to $Q^2 \simeq 1 ~ (GeV/c)^2$; a $CQ$ charge radius of $\simeq 0.45 ~ fm$ has been obtained in this way. Above $Q^2 \simeq 1 ~ (GeV/c)^2$ our parameter-free predictions are found to be in remarkable agreement with the experimental data, showing that soft physics may be dominant up to $Q^2 \sim 10 ~ (GeV/c)^2$. Our results for the longitudinal and transverse helicity amplitudes of the $N - P_{11}(1440)$ transition have been presented. Finally, let us stress that the results presented in this work supersede those of Refs. \cite{CAR_N} and \cite{CAR_Roper}.

\end{document}